\documentclass[12pt,preprint]{aastex}

\begin{document}


\title{Detection of a $45\arcdeg$ Tidal Stream Associated with the
Globular Cluster NGC 5466}

\author{C. J. Grillmair}
\affil{Spitzer Science Center, 1200 E. California Blvd., Pasadena,  CA 91125}
\email{carl@ipac.caltech.edu}
\and
\author{R. Johnson\altaffilmark{1}}
\affil{California State University, Dept. of Physics and Astronomy, 1250 Bellflower Blvd., Long Beach, CA 90840}
\email{rjohnson@ipac.caltech.edu}

\altaffiltext{1}{Spitzer Graduate Student Fellow}

\begin{abstract}

We report on the detection in Sloan Digital Sky Survey data of a
$45\arcdeg$ tidal stream of stars, extending from Bootes to Ursa
Major, which we associate with the halo globular cluster NGC
5466. Using an optimal contrast, matched filter technique, we find a
long, almost linear stellar stream with an average width of
1.4\arcdeg. The stream is an order of magnitude more tenuous than the
stream associated with Palomar 5. The stream's orientation on the sky
is consistent to a greater or lesser extent with existing proper
motion measurements for the cluster.

\end{abstract}



\section{Introduction}

The advent of large-scale, digital sky surveys has enabled a
remarkable increase in our ability to distinguish substructures in our
Galaxy \citep{yann03,maj03,roch04,john05}. Among the first discoveries
in the Sloan Digital Sky Survey (SDSS) data were the remarkably strong
tidal tails of Palomar 5 \citep{oden2001,rock2002, oden2003}, spanning
over $10\arcdeg$ on the sky.  Tidal tails of globular clusters are
particularly interesting from a dynamical standpoint as they are
expected to be very cold \citep{comb99}. This makes them potentially
useful for constraining not only the global mass distribution of the
Galaxy, but also its lumpiness \citep{mura99}.

Globular cluster tidal tails were first discovered in a photographic
survey of 12 Southern halo clusters by
\citet{grill95}. \citet{leon2000} and subsequent workers found similar
evidence for tidal tails in over 30 other Galactic globular
clusters. Once the characteristic, power-law departure at large radius
from a King profile was recognized as a signature of unbound stars,
evidence of tidal tails was detected in globular clusters as far away
as the halo of M31 \citep{grill96}.

NGC 5466 is an interesting globular cluster in many respects. It has a
blue horizontal branch and a large, centrally concentrated
distribution of blue stragglers \citep{nemec87}. On the other hand,
like Pal 5, NGC 5466 is a low metallicity ([Fe/H] = -2.22), low mass,
and low concentration cluster. \citet{pryor91} determined a very low
$(M/L_V)_0 = 1 \pm 0.4$ for this cluster and suggested that it must
have lost a significant fraction of its low mass stars. Combining
these factors with NGC 5466's putative orbit, \citet{gned97} ranked it
among globular clusters most likely to have suffered substantial tidal
stripping over time. Evidence for the existence of unbound stars
around NGC 5466 was first presented by \citet{leh97}. \citet{oden04}
examined the spatial distribution of cluster member candidates within
a degree of NGC 5466 using APM data data. Most recently,
\citet{belo05} used SDSS photometry to discover a tidal stream
extending $2\arcdeg$ from NGC 5466 on either side.

In this paper we examine a much larger region of the SDSS
to search for tidal streams associated with both NGC 5466 and NGC
5272.  We briefly describe our analysis in Section \ref{analysis}. We
discuss our findings for both NGC 5466 and NGC 5272 in Section
\ref{discussion}, and make concluding remarks in Section
\ref{conclusions}.

\section{Data Analysis \label{analysis}}

Data comprising $u',g',r',i'$, and $z'$ measurements and their errors
for $9.7 \times 10^6$ stars in the region $170\arcdeg <$ R.A. $<
230\arcdeg$ and $19\arcdeg < \delta < 51\arcdeg$ were extracted from
the SDSS database using the SDSS CasJobs query system. The data were
analyzed using the matched-filter technique described by
\citet{rock2002}, and the reader is referred to this paper for a
complete description of the method. Briefly, we constructed an
observed color-magnitude density or Hess diagram for NGC 5466 using
stars within $9\arcmin$ of the cluster center. A star count weighting
function was created by dividing this color-magnitude distribution
function by a similarly binned color-magnitude distribution of the
field stars. The Hess diagram for field stars was created using all
stars in rectangular regions to the east and west of NGC 5466, each
region subtending $\sim 150$ square degrees.  The weighting function
was then used on all stars in the field and the weighted star counts
were summed by location in a two dimensional array.

We used all stars with $16 < g' < 22$. As NGC 5466 has $l = 42\arcdeg,
b = 73\arcdeg$, reddening is not expected to be a significant
problem. Nonetheless, variations in foreground reddening across a
large field could introduce corresponding variations in sample
completeness and possibly spurious structures. We therefore
dereddened the SDSS photometry as a function of position on the sky
using the DIRBE/IRAS dust maps of \citet{schleg98}. The applied values of
$E(B-V)$ ranged from 0.006 to 0.06 over a field area of 1446
square degrees.

We optimally filtered the $g' - r'$ and $g' - i'$ star counts
independently as a consistency check. The $u'-g'$ and $g'-z'$ data
were found to be too noisy to contribute significantly to the final
signal-to-noise ratio. As expected, the highest filter weight is given
to turn-off and horizontal branch stars, as these populations lie
blueward of the vast majority of field stars. However, main sequence
stars contribute significantly by virtue of their much larger numbers.

In Figure 1 we show the final, filtered star count distribution,
created by co-adding the weight images generated independently for the
$g'-r'$ and $g'-i'$ color pairs. The image has been smoothed with a
Gaussian kernel of full width $\sigma = 1\arcdeg$. A low-order,
polynomial surface was fitted and subtracted from the image to remove
large scale gradients due to the Galactic disk and bulge. We note that
there are discernible features running parallel to the SDSS scan
directions (or ``stripes''), particularly in the northeastern
quadrant. These are artifacts introduced by variations in seeing and
transparency from one stripe to the next. If we limit our analysis
to stars brighter than $g' = 20$, these features largely disappear.
However, to improve our signal-to-noise ratio we maintain a faint
magnitude cut off of $g' = 22$ and disregard features which mimic the
scan pattern of the survey.

\section{Discussion \label{discussion}}

Extending $\sim 3\arcdeg$ to the southeast of the NGC 5466 and
$2\arcdeg$ to the northwest are fairly strong concentrations of stars
which are obviously connected to the cluster. These are the features
recently reported by \citet{belo05}. Though these tidal arms show
little evidence of the 'S' shape characteristic of weak tidal
stripping, this would be consistent with our current viewing location
very nearly in the plane of the NGC 5466's orbit. Extending $\sim
15\arcdeg$ from NGC 5466 towards the southeast and $\sim 30\arcdeg$ to
the northwest is an almost linear feature which cuts across 
several scan stripes. The feature is $1\arcdeg$-$2\arcdeg$ wide along most of
it's discernible length. This width is similar to that observed in the
tidal tails of Pal 5 and in agreement with the expectation that stars
stripped from globular clusters should have a very low velocity
dispersion. The stream is not associated with the Sagittarius dwarf
debris stream, which runs roughly parallel to the current feature but
is much broader and lies $20\arcdeg$ to the south of the field shown in
Figure 1.

The stream is not a product of our dereddening procedure; careful
examination of the reddening map of \citet{schleg98} shows no
correlation between this feature and the applied reddening
corrections. To be certain, we reran our matched-filter analysis using
the original data uncorrected for reddening. Though less strong, the
feature in Figure 1 remains quite apparent.

The stream is also not due to confusion between galaxies and stars at
faint magnitudes. We have filtered the SDSS Galaxy catalog over the
same field area in a manner identical to that used for objects
classified as stars. There is indeed evidence of confusion in that
there is an obvious concentration of ``galaxies'' within a few
arcminutes of the center of NGC 5466. However, there is no
concentration of galaxies coincident with the extended feature
in Figure 1.

At its southeastern end the stream appears to be truncated by the limits
of the available data. On the northwestern end, the stream becomes
indiscernible westward of RA = 180\arcdeg. Based on the proper motion
\citep{oden97} and assumed orbit of the cluster (see below), the
northwestern end of the current tidal stream should be roughly a
factor of two further away from us than the cluster itself. In an
attempt the trace the stream still farther westwards, we reapplied the
matched filter after shifting the NGC 5466 color-magnitude sequence
fainter by 1.5 magnitudes. However, no evidence for a continuing
stream could be detected using the present analysis. While it is
conceivable that we are seeing the physical end of the stream (and
therefore the very first stars to be stripped from NGC 5466) it seems
more likely that our failure to trace the stream any further simply
reflects increasing contamination by field stars at fainter magnitudes
and the reduced power of the optimal filter.

In Figure 2 we show the locations of $\sim 150$ stars with colors and
magnitudes which would put them on the blue end ($g' - r' < 0$) of NGC
5466's horizontal branch. While the statistics are small and there are
interesting groupings here and there across the field, these blue
stars show a marked tendency to lie along the tidal stream.  Measuring
over a radius range of $1\arcdeg$ to $16\arcdeg$ from the cluster
these HB stars are between 2 and 3 times more likely to be found
within $1.5\arcdeg$ of the center of the tidal stream.

There are seven lines of reasoning that indicate that the feature in
Figure 1 is indeed a tidal stream associated with NGC 5466. 1) We know
from the work of \citet{belo05} that NGC 5466 possesses tidal tails
extending at least $4\arcdeg$ on the sky. 2) Our optimal filter
technique recovers the strong, $2\arcdeg$ tidal tails found by
\citet{belo05}.  3) The projection of the $45\arcdeg$-long feature
passes within $30\arcmin$ of the center of NGC 5466. 4) The feature
disappears if the Hess diagrams used to create the matched filters are
shifted either blueward or redward of the observed color-magnitude
sequence of NGC 5466. 5) The apparent width of the feature is in
accord with expectations for globular cluster tidal tails (e.g. Pal
5).  6) Though much noisier, the distribution of candidate horizontal
branch stars also shows a tendency to concentrate along the stream
defined by main sequence/RGB stars.  7) Given the expectation that the
stream traces the orbital path of the cluster, its orientation is
consistent with the previously derived space motion of the cluster
(see below).

Integrating the weighted star counts along each stream over a width of
$1.4\arcdeg$ and beyond $2\arcdeg$ from NGC 5466 itself, we find the
total number of stars in the discernible streams to be $607 \pm
50$. Based on a comparison between the observed SDSS luminosity
function of NGC 5466 (measured outside the saturated central region)
and that of other well studied globulars, we estimate that between two
thirds and three quarters of the cluster stars are missing from the
SDSS field sample due to incompleteness at $g' > 21$. The apparent
stream must therefore contain at least $\sim 2 \times 10^3$ main
sequence stars. If the leading and trailing arms are assumed to be
symmetric, then we might expect another thousand stars in the region
not covered by the SDSS DR4. In total this represents only about 3\%
of the current mass of the cluster. However, if the luminosity
function of the tails is heavily biased towards low mass (i.e. the
missing low mass stars postulated by \citet{pryor91}), then both the
number of stars and the cluster mass fraction resident in the tidal
tails may be much higher.

The surface density of stars fluctuates considerably along each
stream. For stars with $g' < 22$ the average surface density is $10
\pm 2$ stars per square degree, with several peaks along the streams
ranging from 20 to over 30 stars per square degree. This may be
compared with a total surface density of 7600 field stars per degree
squared for all stars with $g' < 22$ and illustrates the remarkable
power of matched filtering. The tidal tails of NGC 5466 are evidently
much more tenuous than those of Pal 5, for which \citet{oden2003}
found surface densities typically in excess of 100 stars per square
degree. However, the relative magnitudes of the surface density
fluctuations along the length of the stream are quite similar.

\subsection{On the Orbit of NGC 5466} \label{orbit}

Assuming globular cluster tidal streams follow the orbits of their
parent clusters \citep{oden2003}, we can use the observed orientation
of the newly discovered stream to better constrain the orbit of NGC
5466. We use the Galactic model of \citet{allen91} for orbit
integration. Using the Lick-based proper motion measurements of
\citet{bros91} we find a projected orbit as shown by the solid line in
Figure 2. Using \citet{oden97}'s Hipparcos-derived proper motions we
find the orbit projection shown by the dotted line. The uncertainties
on the two measurements are similar, and the plate-based measurements
are in agreement with Hipparcos measurements at the $\approx
1.5\sigma$ level. The orbit projections diverge from one another by
$\approx 20\arcdeg$, and the tidal stream clearly favors the
ground-based proper motion measurement. 

Even using the \citet{bros91} proper motion measurement, the match
between the projected orbit and the tidal stream is not completely
satisfactory. The northwestern stream does not lie perfectly along any
computed orbit and, indeed, appears to depart in places from a smooth
curve. This is due in part to small number statistics, but could also
be due to 1) irregularities in the potential of the Galactic halo, 2)
a recent weak encounter between stream stars and a substantial mass
concentration in the disk (e.g. a large molecular cloud), or 3)
confusion. If we ignore the apparent enhancements westward of RA =
$190\arcdeg$, the stream becomes somewhat less meandering and more
closely resembles the computed orbits. If we set (U, V, W) = (290,
-240, 225 km s$^{-1}$), we arrive at the dash-dot line in Figure
2. Since our vantage point is very nearly within the plane of NGC
5466's orbit, we would not expect to see a significant 
offset between the computed orbit and the leading and trailing
tails. The U,V,W velocities required to match the stream depart from
the proper motion measurement of \citet{bros91} at the $0.5\sigma$
level, and from the Hipparcos measurement at the $2.2\sigma$ level.

All reasonable orbits predict a total length of $\simeq 15$ kpc for
the northwestern arm, and $\simeq 5$ kpc for the southeastern. At the
end of the northwestern arm, the orbital path is inclined to
our line of sight by some $35\arcdeg$.

\subsection{NGC 5272}

Similar matched filtering procedures were applied to the present
dataset to search for tidal streams associated with NGC 5272. Based on
a color-magnitude sequence which has a somewhat bluer RGB and a
brighter turnoff than that of NGC 5466, we expected to be considerably
more sensitive to any tidal streams associated with this
cluster. However, no evidence for tidal streams could be found. This
is in qualitative agreement with the results of \citet{gned97}, who
estimated a probability of disruption for NGC 5272 roughly sixty times
lower than that for NGC 5466.

\section{Conclusions \label{conclusions}}

Applying optimal contrast filtering techniques to SDSS data, we have
detected a tenuous stream of stars roughly $45\arcdeg$ long on the
sky. Based on several pieces of evidence, we associate this stream with
the halo globular cluster NGC 5466.  

Verification of the stream and its association with NGC 5466 will
require radial velocity measurements of individual stars along its
length. Based purely on SDSS colors, there are $\sim 350$ candidate
stream stars along the length of the tail with $g' < 19.2$ (i.e. above
the subgiant branch). Of these, we expect $\sim 20$ stars to be {\it
bona fide} orphans of NGC 5466.  Once vetted, these stars will become
prime targets for the Space Interferometry Mission, whose proper
motion measurements will enable very much stronger constraints to be
placed on both the orbit of the cluster and on the potential field of
the Galaxy.

\acknowledgments

Funding for the creation and distribution of the SDSS Archive has been
provided by the Alfred P. Sloan Foundation, the Participating
Institutions, the National Aeronautics and Space Administration, the
National Science Foundation, the U.S. Department of Energy, the
Japanese Monbukagakusho, and the Max Planck Society. The SDSS web site
is http://www.sdss.org/.

We are also very grateful to an anonymous referee for several
suggestions which greatly improved the manuscript.

{\it Facilities:} \facility{Sloan}.

\clearpage

\begin{figure}
\epsscale{.80}
\plotone{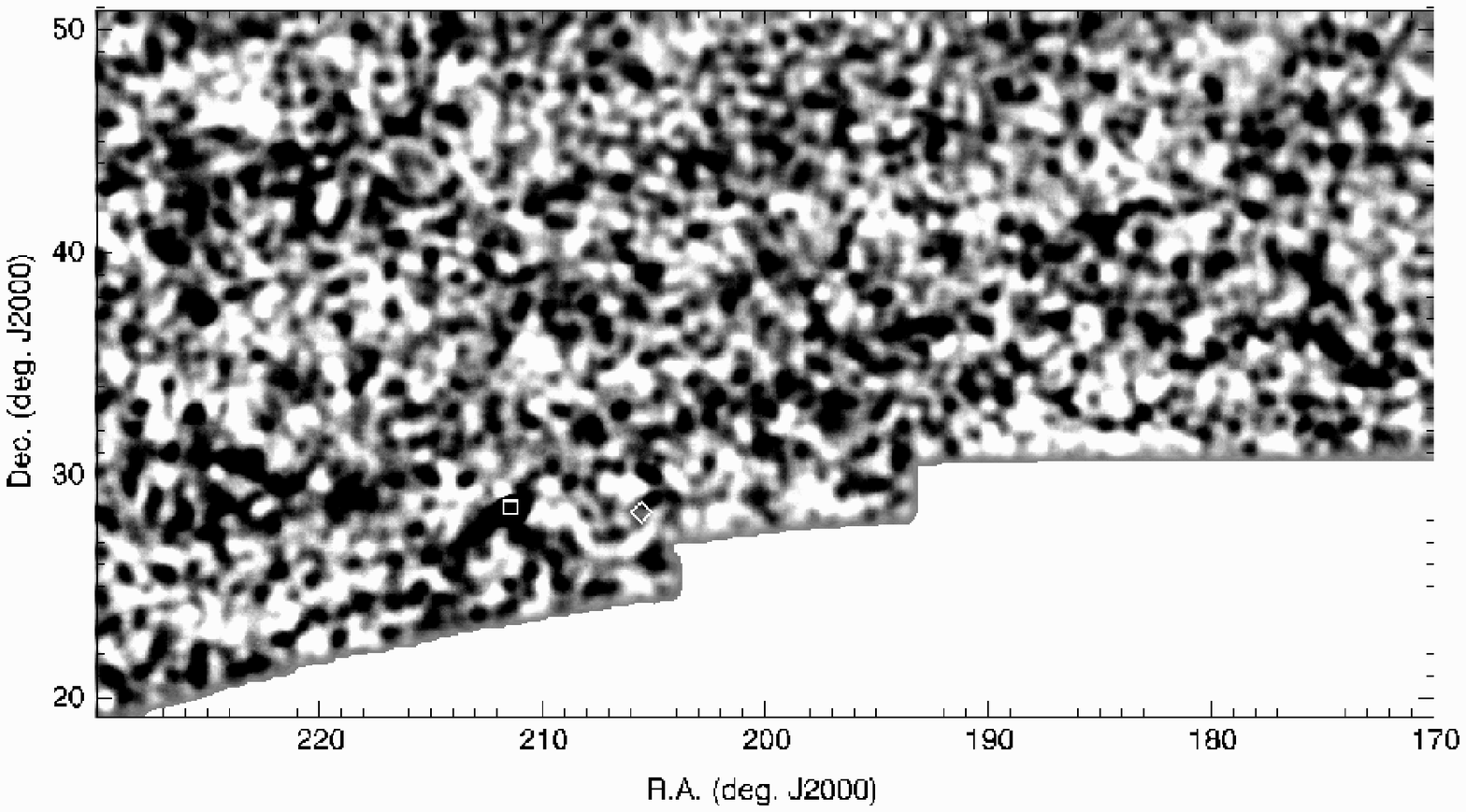}
\caption{Smoothed, summed weight image of the SDSS field after
subtraction of a low-order surface fit. Darker areas indicate higher
surface densities. The image is the sum of weight images generated
independently using the $g' - r'$ and $g' - i'$ color pairs. NGC 5466
is indicated by the open square at R.A., dec = (211.36, +28.54) while
the location of NGC 5272 (R.A., dec = 205.55, +28.38) is shown by the
open diamond. The weight image has been smoothed with a Gaussian
kernel of full width 1\arcdeg. The irregular southern border is
defined by the limits of SDSS Data Release 4. The faint, parallel
features in the northeastern corner trace the edges of individual SDSS
scans and are presumably due to variations in sensitivity and
completeness at faint magnitude levels. The putative tidal stream of
NGC 5466 extends from southeastern corner of the image to roughly
R.A., dec = (180, 42). \label{fig1}}
\end{figure}

\clearpage

\begin{figure}
\epsscale{1.0}
\plotone{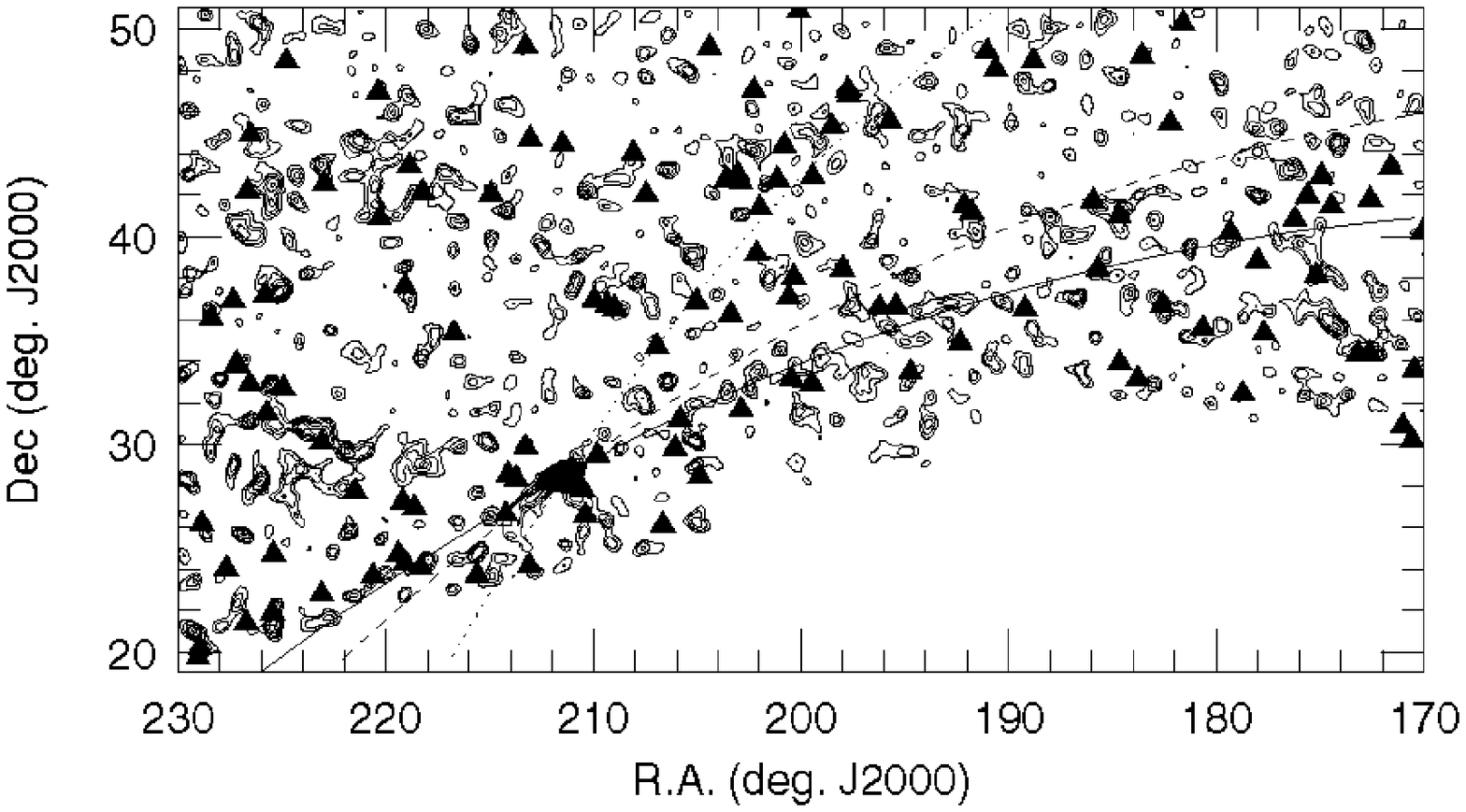}
\caption{Orbit integrations for NGC 5466. Contours are taken from the
weight image in Figure 1 and represent the $1, 2, 3, ... n \sigma$
levels. The solid line shows an orbit projection based on proper
motion measurements of \citet{bros91}, and the dotted line shows a
similar projection using the Hipparcos-derived proper motion
\citep{oden97}. The dash-dot line uses (U,V,W) = (290, -240, 225) km
s$^{-1}$.  The filled triangles show the locations of stars with
colors and magnitudes consistent with those of the bluest horizontal
branch stars in NGC 5466.}
\end{figure}

\end{document}